\documentclass{jfm}
\usepackage{graphicx}
\usepackage{newtxtext}
\usepackage{newtxmath}
\usepackage{natbib}
\usepackage{hyperref}
\hypersetup{
    colorlinks = true,
    urlcolor   = blue,
    citecolor  = black,
}

\newcommand{\RomanNumeralCaps}[1]
\linenumbers

\newcommand{\diff}{\mathrm{d}}
\newcommand{\DR}{\mathrm{DR}}

\title{Direct numerical simulation of drag reduction in rotating pipe flow up to $\Rey_\tau \approx 3000$}

\author{Maochao Xiao,
  Alessandro Ceci,
  Paolo Orlandi,
  \and Sergio Pirozzoli
  \corresp{\email{sergio.pirozzoli@uniroma1.it}}}

\affiliation{Dipartimento di Ingegneria Meccanica e Aerospaziale, Sapienza Università di Roma, Via Eudossiana 18, 00184 Roma, Italy}

\begin{document}
\maketitle

\begin{abstract}
Direct numerical simulations (DNS) of rotating pipe flows up to $\Rey_\tau \approx 3000$ are carried out to investigate drag reduction effects associated with axial rotation, extending previous studies carried out at a modest Reynolds number~\citep{Orlandi1997a, Orlandi2000}. The results show that the drag reduction, which we theoretically show to be equivalent to net power saving assuming no mechanical losses, monotonically increases as either the Reynolds number or the rotation number increases, proportionally to the inner-scaled rotational speed. Net drag reduction up to about $70\%$ is observed, while being far from flow relaminarisation. Scaling laws for the mean axial and azimuthal velocity are proposed, from which a predictive formula for the friction factor is derived. The formula can correctly represent the dependency of the friction factor on the Reynolds and rotation numbers, maintaining good accuracy for low-to-moderate rotation numbers. Examination of the turbulent structures highlights the role of rotation in widening and elongating the small-scale streaks, with subsequent suppression of sweeps and ejections. In the core part of the flow, clear weakening of large-scale turbulent motions is observed at high Reynolds numbers, with subsequent suppression of the outer-layer peak in the pre-multiplied spectra. The Fukagata-Iwamoto-Kasagi decomposition indicates that, consistent with a theoretically derived formula, the outer layer yields the largest contribution to drag reduction at increasingly high Reynolds numbers. In contrast, both the inner and the outer layers contribute to drag reduction as the rotation number increases.
\end{abstract}


\section{Introduction}\label{sec:intro}

Turbulent flow in circular pipes has always attracted the interest of scientists, owing to its prominent importance in engineering practice and because of the beautiful simplicity of the setup. The study of flows in circular pipes holds particular significance not only for its fundamental insights into fluid dynamics but also for its practical applications, especially in industries such as oil and gas transportation. Understanding the behaviour of turbulent flow in pipelines is crucial for optimizing transport efficiency, ensuring structural integrity, and minimizing energy consumption. As such, research into flows in circular pipes continues to be an active area of study, driving innovations in both theoretical understanding and engineering practice.

Specifically, the research on drag reduction techniques in pipe flow holds significant importance due to its potential for reducing energy consumption in the transportation of gases or liquids, thereby leading to decreased pollutant emissions into the atmosphere. 
The potential of axial rotation for drag reduction in turbulent pipe flow was first shown by \citet{White1963}. In those experiments, dye was injected from a hypodermic tube into the central core of the flow in a pipe, and then diffused radially outwards to fill the whole pipe. When rotation was imposed on the pipe wall, the dye moved along the central core of the pipe without as much radial diffusion. The rate of rotation is typically measured in terms of the rotation number, $N = \Omega R / u_b$, where $\Omega$ is the rotational speed, $R$ is the pipe radius, and $u_b$ is bulk velocity. The experiments of \citet{Kikuyama1983} showed that as $N$ increases, the wall friction decreases, with the mean axial velocity profiles approaching the parabolic Poiseuille solution. These findings were subsequently confirmed from the direct numerical simulations (DNS) of \citet{Orlandi1997a}. Those authors mainly attributed the mechanism of drag reduction to the stabilising effect of the radially growing centrifugal force (or angular momentum). Indeed, according to the Rayleigh criterion~\citep{Reich1989,Drazin2004}, the radially growing pressure gradients impede the radial motion of fluid particles. \citet{Orlandi1997a} further pointed out that axial rotation induces a long columnar structure in the core region which transports the streamwise vortical structures away from the wall, while tilting and widening the near-wall streaks. Also, wall rotation disrupts the symmetry between right- and left-handed helical structures, and the resulting high helicity density yields less dissipation, with incurred drag reduction~\citep{Orlandi1997b}. \citet{Zhang2022} showed that the sweep events are systematically suppressed by rotation, which further impedes the formation of hairpin structures. However, they found that rotation amplifies the azimuthal and radial velocity variances, and that the pressure-strain and Coriolis production terms become the leading terms in the budget of these two Reynolds normal stresses.

\citet{Davis2019} reported that the drag reduction effect increases with the bulk Reynolds number, $\Rey_b = u_b D/ \nu$, where $D=2 R$ is the pipe diameter, and $\nu$ is kinematic viscosity. However, their highest friction Reynolds number, $\Rey_{\tau} = u_\tau R/\nu$ (with $u_\tau=(\tau_w/\rho)^{1/2}$ the friction velocity), was about 540, at which extrapolation to real-world scenarios of fully developed turbulent flow is far from clear, and well less than achieved in DNS of non-rotating pipe flow~\citep{Pirozzoli2021b}. 

Other techniques for drag reduction in turbulent pipe flow have been proposed based 
on more complex wall actuation. \citet{Quadrio2000} studied turbulent flow in a 
circular pipe oscillating around its longitudinal axis through DNS and found that 
the maximum amount of drag reduction achievable with appropriate oscillations of 
the pipe wall is on the order of 40\%, comparable to what was found in planar geometries. 
They observed that the transverse shear layer induced by wall oscillation induces substantial 
modifications to the turbulence statistics in the near-wall region, indicating a strong effect 
on the vortical structures. \citet{Auteri2010} assessed the practical use of travelling waves 
of spanwise velocity, as suggested by \citet{Quadrio2009} and further analyzed by \citet{Gatti2016}, 
to achieve drag reduction in pipe flow. In their experiments, the pipe wall was subdivided into 
thin slabs that could rotate independently in the azimuthal direction, confirming the possibility of 
achieving drag reduction of up to 33\%.

In the present study, we leverage DNS data to explore drag reduction in turbulent pipe flow up to $\Rey_\tau \approx 3000$ through the use of steady axial rotation. The goal is to verify and quantify the drag reduction effects, and shed light on the underlying physical mechanisms.

\section{The numerical dataset}\label{sec:method}

A second-order finite-difference solver is used to solve the incompressible Navier-Stokes equations in cylindrical coordinates~\citep{Orlandi1997a,Pirozzoli2021b}. The current DNS pertain to fully developed turbulent flow, with periodic conditions along the axial direction. The DNS are carried out in a frame of reference rotating with the pipe, which has the advantage of allowing for larger time steps as compared to the inertial frame. Coriolis forces are then added as $-2\Omega u_r$ and $2\Omega u_{\theta}$ to the azimuthal and radial momentum equations, where $u_\theta$ and $u_r$ are the velocity components in the azimuthal and radial directions.
The simulations were conducted under conditions of constant mass flow rate. From now on, the subscript $0$ indicates non-rotating cases, and the superscript $*$ 
is used to denote normalisation with wall units. 
Normalisation with wall units based on the non-rotating case is also used to better highlight the effects of pipe rotation, which we indicate with the $+$ superscript.
We performed DNS for various bulk Reynolds numbers, namely $\Rey_b = 5300, 17000, 44000, 82500$, and $133000$, corresponding to friction Reynolds numbers of $\Rey_{\tau,0} = 180$, $495$, $1137$, $1979$ and $3020$, in the absence of rotation. For each Reynolds number, we have considered various rotation numbers, namely $N=0$, $0.25$, $0.5$, $1.0$ and $2.0$ and $4.0$. A list including the flow parameters for all the simulations is provided in table~\ref{tab:params}. The pipe length is set to $L=15R$, and the mesh resolution for each of the Reynolds numbers is decided based on the non-rotating cases, which exhibit the highest wall friction. Specifically, the grid spacing is $\Delta z^+ \approx 10$, and $R^+ \Delta \theta \approx 4$, along the axial and azimuthal directions, respectively. In the radial direction, the grid spacing is uniform up to $y^+ \approx 5$, and then proportional to the local Kolmogorov length scale ($\eta^+ \approx 0.8(y^+)^{1/4}$) in the outer layer. About thirty grid points are allocated for $y^+ \le 40$, with the first grid point located at $y^+ \approx 0.05$. Additional details can be found in previous publications~\citep{Pirozzoli2021a}. The sensitivity of the results to pipe length and mesh resolution is analyzed in Appendix~\ref{app:A}. Hereafter, capital letters will be used to denote flow properties averaged in the homogeneous spatial directions and in time, brackets to denote the averaging operator, and lower-case letters to denote fluctuations from the mean.

\begin{table}
  \begin{center}
\def~{\hphantom{0}}
  \begin{tabular}{lcccccccc}
    $\Rey_b$ & $\Rey_{\tau,0}$	&$N$       &Mesh ($N_\theta \times N_r \times N_z$)  &$\Rey_{\tau} $    & $N^+$		& $\lambda \times 10^3$    &  $\DR$ (\%)   			&  \#ETT   \\[3pt]
	5300	  &180.10			&0	       &256 $\times$ 55 $\times$ 256		      &180.10		  	&0.0	&$37.469 \pm 0.24\%$	   &$0.00  \pm 0.00\%$ 		&74.76\\
	5300	  &180.10			&0.25	   &256 $\times$ 55 $\times$ 256		      &174.69		  	&3.7	&$34.748 \pm 0.35\%$	   &$7.26 	\pm 5.42\%$ 	&72.51\\
	5300	  &180.10			&0.5	   &256 $\times$ 55 $\times$ 256		      &168.27		  	&7.4	&$32.256 \pm 0.52\%$	   &$13.80	\pm 3.54\%$ 	&10.64\\
	5300	  &180.10			&1.0	   &256 $\times$ 55 $\times$ 256		      &168.62		  	&14.7	&$32.388 \pm 0.45\%$	   &$12.32	\pm 3.25\%$ 	&70.60\\
	5300	  &180.10			&2.0	   &256 $\times$ 55 $\times$ 256		      &168.43		  	&29.4	&$32.317 \pm 0.58\%$	   &$12.52	\pm 3.94\%$ 	&70.02\\
	5300	  &180.10			&4.0	   &256 $\times$ 55 $\times$ 256		      &167.26		  	&58.9	&$31.869 \pm 0.76\%$	   &$13.73	\pm 4.54\%$ 	&70.34\\
	17000	  &495.29			&0	       &768 $\times$ 96 $\times$ 768		      &495.29		  	&0.0	&$27.160 \pm 0.09\%$	   &$0.00	\pm 0.00\%$ 	&35.00\\
	17000	  &495.29			&0.25	   &768 $\times$ 96 $\times$ 768		      &481.12		  	&4.3	&$25.624 \pm 0.10\%$	   &$5.80	\pm 2.24\%$ 	&42.51\\
	17000	  &495.29			&0.5	   &768 $\times$ 96 $\times$ 768		      &452.65		  	&8.6	&$22.681 \pm 0.16\%$	   &$16.62	\pm 0.93\%$ 	&26.68\\
	17000	  &495.29			&1.0	   &768 $\times$ 96 $\times$ 768		      &411.53		  	&17.2	&$18.747 \pm 0.29\%$	   &$31.08	\pm 0.68\%$ 	&24.26\\
	17000	  &495.29			&2.0	   &768 $\times$ 96 $\times$ 768		      &396.91		  	&34.3	&$17.439 \pm 0.44\%$	   &$35.89	\pm 0.81\%$ 	&23.39\\
	17000	  &495.29			&4.0	   &768 $\times$ 96 $\times$ 768		      &376.59		  	&68.6	&$15.700 \pm 0.37\%$	   &$42.28	\pm 0.52\%$ 	&22.20\\
	44000	  &1136.59			&0	       &1792 $\times$ 270 $\times$ 1792	          &1136.59	      	&0.0	&$21.352 \pm 0.10\%$	   &$0.00	\pm 0.00\%$ 	&25.86\\
	44000	  &1136.59			&0.25	   &1792 $\times$ 164 $\times$ 1792	          &1087.91	      	&4.8	&$19.562 \pm 0.14\%$	   &$8.38	\pm 1.88\%$ 	&24.78\\
	44000	  &1136.59			&0.5	   &1792 $\times$ 164 $\times$ 1792	          &1013.63	      	&9.7	&$16.982 \pm 0.08\%$	   &$20.47	\pm 0.50\%$ 	&34.44\\
	44000	  &1136.59			&1.0	   &1792 $\times$ 164 $\times$ 1792	          &892.25	      	&19.4	&$13.158 \pm 0.14\%$	   &$38.37	\pm 0.28\%$ 	&24.35\\
	44000	  &1136.59			&2.0	   &1792 $\times$ 164 $\times$ 1792	          &808.56	      	&38.7	&$10.806 \pm 0.49\%$	   &$49.39	\pm 0.51\%$ 	&13.51\\
	44000	  &1136.59			&4.0	   &1792 $\times$ 164 $\times$ 1792	          &755.78	      	&77.4	&$9.441  \pm 0.54\%$       &$55.78	\pm 0.44\%$ 	&29.90\\
        44000*    &1136.59			&4.0	   &1792 $\times$ 164 $\times$ 3584	          &767.70	      	&77.4 	&$9.741  \pm 0.26\%$       &$54.38	\pm 0.23\%$ 	&25.49\\
        44000**   &1136.59			&4.0	   &1792 $\times$ 328 $\times$ 1792	          &742.94	      	&77.4 	&$9.123  \pm 0.00\%$       &$54.38	\pm 0.00\%$ 	&50.60\\
	82500	  &1979.32			&0	       &3072 $\times$ 399 $\times$ 3072	          &1979.32	      	&0.0	&$18.420 \pm 0.18\%$	   &$0.00	\pm 0.00\%$ 	&18.62\\
	82500	  &1979.32			&0.25	   &3072 $\times$ 243 $\times$ 3072	          &1861.11	      	&5.2	&$16.285 \pm 0.14\%$	   &$8.38	\pm 1.74\%$ 	&24.45\\
	82500	  &1979.32			&0.5	   &3072 $\times$ 243 $\times$ 3072	          &1723.57	      	&10.4	&$13.967 \pm 0.25\%$	   &$20.47	\pm 0.97\%$ 	&23.67\\
	82500	  &1979.32			&1.0	   &3072 $\times$ 243 $\times$ 3072	          &1498.29	      	&20.8	&$10.554 \pm 0.42\%$	   &$38.37	\pm 0.61\%$ 	&28.33\\
	82500	  &1979.32			&2.0	   &3072 $\times$ 243 $\times$ 3072	          &1316.85	      	&41.7	&$8.153  \pm 1.10\%$       &$55.74	\pm 0.89\%$ 	&19.32\\
	82500	  &1979.32			&4.0	   &3072 $\times$ 243 $\times$ 3072	          &1221.82	      	&83.4	&$7.019  \pm 1.42\%$       &$55.78	\pm 0.88\%$ 	&9.39\\
	133000	  &3020.16			&0	       &4608 $\times$ 327 $\times$ 4608	          &3020.16	      	&0.0	&$16.501 \pm 0.26\%$	   &$0.00	\pm 0.00\%$ 	&6.97\\
	133000	  &3020.16			&0.25	   &4608 $\times$ 327 $\times$ 4608	          &2820.43	      	&5.5	&$14.390 \pm 0.37\%$	   &$13.44	\pm 3.08\%$ 	&7.12\\
	133000	  &3020.16			&0.5	   &4608 $\times$ 327 $\times$ 4608	          &2586.30	      	&11.0	&$12.100 \pm 0.27\%$	   &$27.21	\pm 1.03\%$ 	&13.51\\
	133000	  &3020.16			&1.0	   &4608 $\times$ 327 $\times$ 4608	          &2275.20	      	&22.0	&$9.364  \pm 0.54\%$       &$43.67	\pm 0.79\%$ 	&5.97\\
	133000	  &3020.16			&2.0	   &4608 $\times$ 327 $\times$ 4608	          &1882.89	      	&44.0	&$6.413  \pm 0.30\%$       &$61.42	\pm 0.25\%$ 	&6.46\\
	133000	  &3020.16			&4.0	   &4608 $\times$ 327 $\times$ 4608	          &1699.39	      	&88.1 	&$5.224  \pm 0.73\%$       &$68.57	\pm 0.36\%$ 	&5.66\\
  \end{tabular}
	  \caption{Flow parameters for DNS. $\Rey_b = 2 R u_b / \nu$ is the bulk Reynolds number, $\Rey_{\tau} = u_\tau R/\nu$ is the friction Reynolds number, $\Rey_{\tau,0}$ is the friction Reynolds number for the non-rotating case, $N = \Omega R / u_b$ is the rotation number, $N_\theta$, $N_r$ and $N_z$ are the number of grid points along the azimuthal, radial and axial directions, respectively, $N^+=\Omega R/u_{\tau,0}$ is the inner-scaled rotational speed; $\lambda$ is the friction factor, $\DR$ is the percentage of drag reduction and \#ETT is the time-averaging interval, expressed in terms of the eddy turnover time, $R/u_\tau$. The symbol $*$ denotes DNS with pipe length $L = 30R$, and $**$ denotes DNS with doubled resolution in the radial direction.} The standard uncertainty of the friction factor is estimated using a modified batch means method~\citep{Russo2017}, and the uncertainty of drag reduction is evaluated by propagating the standard uncertainties of the friction factors.
  \label{tab:params}
  \end{center}
\end{table}

\section{Results}\label{sec:results}

\subsection{Drag reduction}

\begin{figure}
\centering
\includegraphics[width=\linewidth]{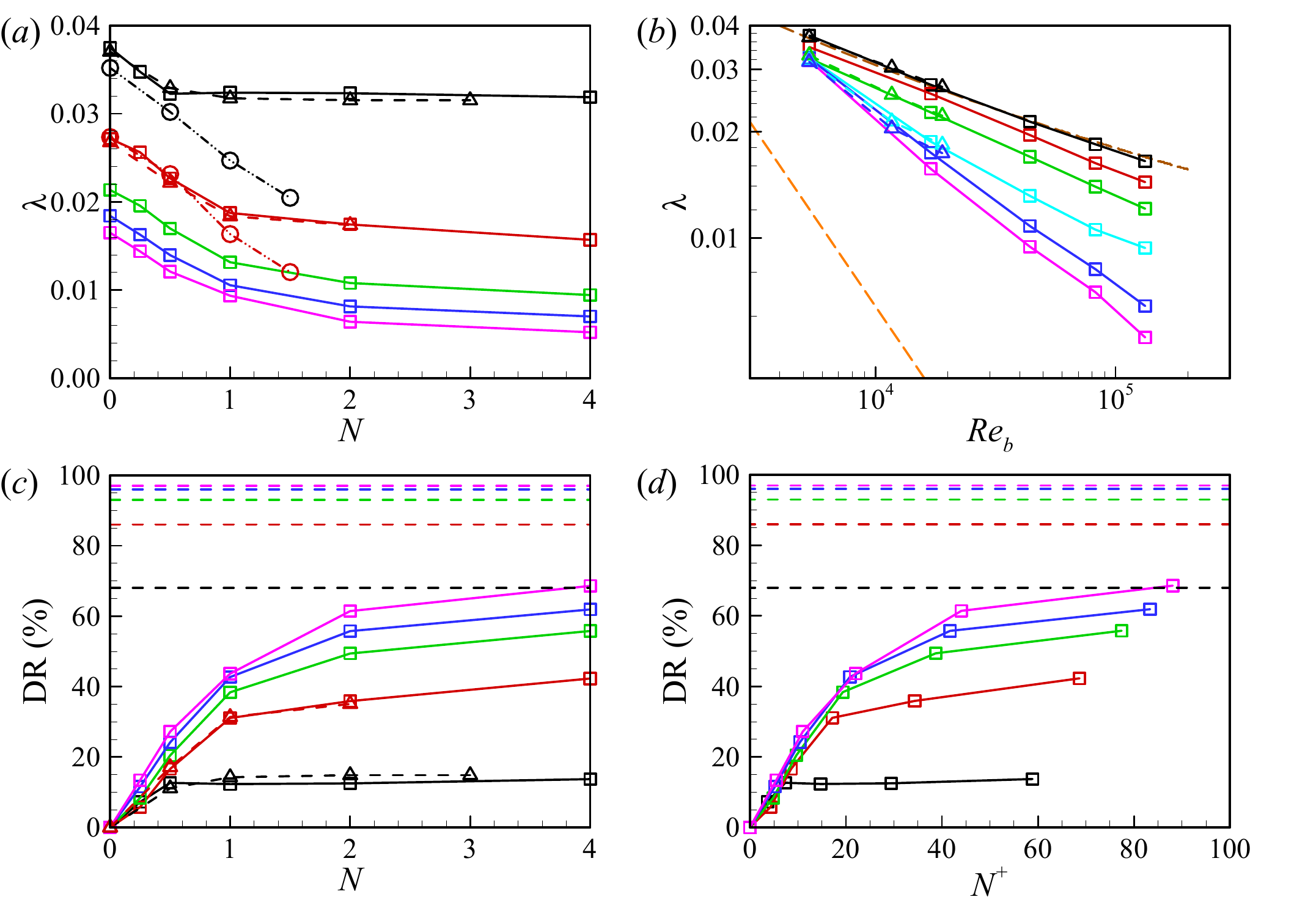}
	\caption{Global flow properties: (a) friction factor as a function of $N$; (b) friction factor as a function of $\Rey_b$; (c) drag reduction coefficient as a function of $N$; (d) drag reduction coefficient as a function of $N^+$.} In (a), (c) and (d), we show data for $\Rey_{\tau,0} = 180$ (black), $495$ (red), $1137$ (green), $1979$ (blue) and $3020$ (pink); in (b) we show $N=0$ (black), $0.25$ (red), $0.5$ (green), $1.0$ (cyan), $2.0$ (blue) and $4.0$ (pink). In (b), the brown dashed line denotes Prandtl's friction law~\citep{Pirozzoli2021b}, and the orange dashed line denotes the laminar friction law $\lambda = 64/\Rey_b$. In (c) and (d), the horizontal dashed lines denote the laminar limits of drag reduction for various Reynolds numbers. The triangle symbols denote DNS data at $\Rey_{\tau,0} \approx 180$ and $540$~\citep{Davis2019}, and the circles denote experiments at $\Rey_{\tau,0} \approx 240$ and $570$~\citep{Kikuyama1983}.
\label{fig:drag}
\end{figure}

Before delving into the analysis of drag reduction, we compare our results with previous studies.
Figure~\ref{fig:drag} displays the friction factor results. For the non-rotating cases, the DNS results exhibit minor derivations from Prandtl's friction law~\citep{Pirozzoli2021b}. 
For the rotating cases, the DNS results for $\Rey_{\tau,0} = 180$ and $\Rey_{\tau,0} = 495$ agree well with the DNS data of \citet{Davis2019}. However, all DNS results exhibit large discrepancies from the experimental data of \citet{Kikuyama1983} at high rotation number, even accounting for differences in the Reynolds number ($\Rey_{\tau,0} = 240$ and $570$ for those authors). \citet{Orlandi1997a} attributed discrepancies to the influence of the entrance conditions, and possible difficulties in achieving a constant pressure gradient in the experimental setup. We note that differences could also be related to limited pipe length, which might prevent the achievement of a fully developed state in spatially developing flows. Indeed, as shown in Appendix~\ref{app:A}, see figure~\ref{fig:dpdz}), the time interval needed to achieve a fully developed state is proportional proportional to $N$, for given $\Rey_b$. 
The drag reduction coefficient $\DR$ is here defined as
\begin{equation}\label{eq:dr}
  \DR = 1 - \frac{C_f}{C_{f,0}} = 1 - \frac{\lambda}{\lambda_0},
\end{equation}
where $\lambda = 4 C_{f}$ is the friction factor, and again the subscript $0$ indicates non-rotating cases. The average power expenditure to sustain wall rotation is
\begin{equation}\label{eq:power}
	P = 2\pi \Omega L R^2 \left. \tau _{r\theta } \right|_{r = R} = 2\pi \mu {\Omega }L R^3 {\left. \frac{\diff}{\diff r} \left( \frac{V_\theta }{r} \right) \right|_{r = R}},
\end{equation}
where $\tau_{r\theta}$ is the tangential viscous shear stress, and $V_{\theta } = U_{\theta} + \Omega r$
is the mean azimuthal velocity in the inertial frame of reference. 
As formally shown below, mean momentum balance under the assumption of statistically steady flow implies that $\diff (V_\theta/r)/ \diff r$ is zero at the wall, i.e., there is no mean azimuthal friction. 
Hence, whereas energy must be spent to maintain the mass flow rate and to rotate the pipe during the initial transient, no energy must be spent to sustain wall rotation once statistically steady conditions are achieved, let alone mechanical losses in the actuation system. As a consequence, the drag reduction coefficient~\eqref{eq:dr} is identical to the net power saving, which establishes an important difference and advantage from other types of wall manipulation requiring additional actuation energy~\citep[e.g.][]{Ricco2021}.

Figure~\ref{fig:drag}(c) shows the variation of the drag reduction coefficient with the rotation number at constant values of $\Rey_b$ (or $\Rey_{\tau,0}$). Pipe rotation consistently leads to drag reduction, 
and monotonic increase with both $N$ and $\Rey_{\tau,0}$ is observed, in agreement with previous studies~\citep{Davis2019}. 
At $N=4.0$ and $\Rey_{\tau,0} = 3020$, drag reduction is as high as 69\%, which clearly highlights the potential 
of rotation for curtailing energy consumption in high-$\Rey$ internal flows, such as fluid transportation 
in large-diameter pipelines, whose typical Reynolds number is $\Rey_{\tau,0} = 10^5-10^6$~\citep{Hultmark2012}. 
This is another important difference from other passive and active drag reduction strategies, 
for which the drag reduction effect typically decreases with the Reynolds number~\citep{Gatti2016, Ricco2021}. We note that full relaminarisation would result in a drag reduction of about $97\%$ at $\Rey_{\tau,0} = 3020$, so at even $N=4.0$, the flow remains distant from the laminar state.
Figure~\ref{fig:drag}(d) displays the drag reduction coefficient as a function of the inner-scaled wall rotational speed based on the non-rotating friction velocity,
$N^+ = \Omega R /u_{\tau,0}$,
which again confirms increased drag reduction with both $N^+$ and $\Rey_{\tau,0}$.
Notably, the drag reduction profiles tend to be much more universal when expressed as a function of $N^+$, with  
deviations from a common trend occurring at higher and higher $N^+$ as $\Rey_{\tau,0}$ increases. 
This suggests that the proper parameter to quantify drag reduction effects could be the ratio of the pipe peripheral velocity to the friction velocity, which is the typical scale for wall turbulence, in line with what is found in drag reduction studies based on the use of 
oscillating walls and traveling waves~\citep{Quadrio2009,Touber2012}. 

\begin{figure}
  \centering
  \includegraphics[width=\linewidth]{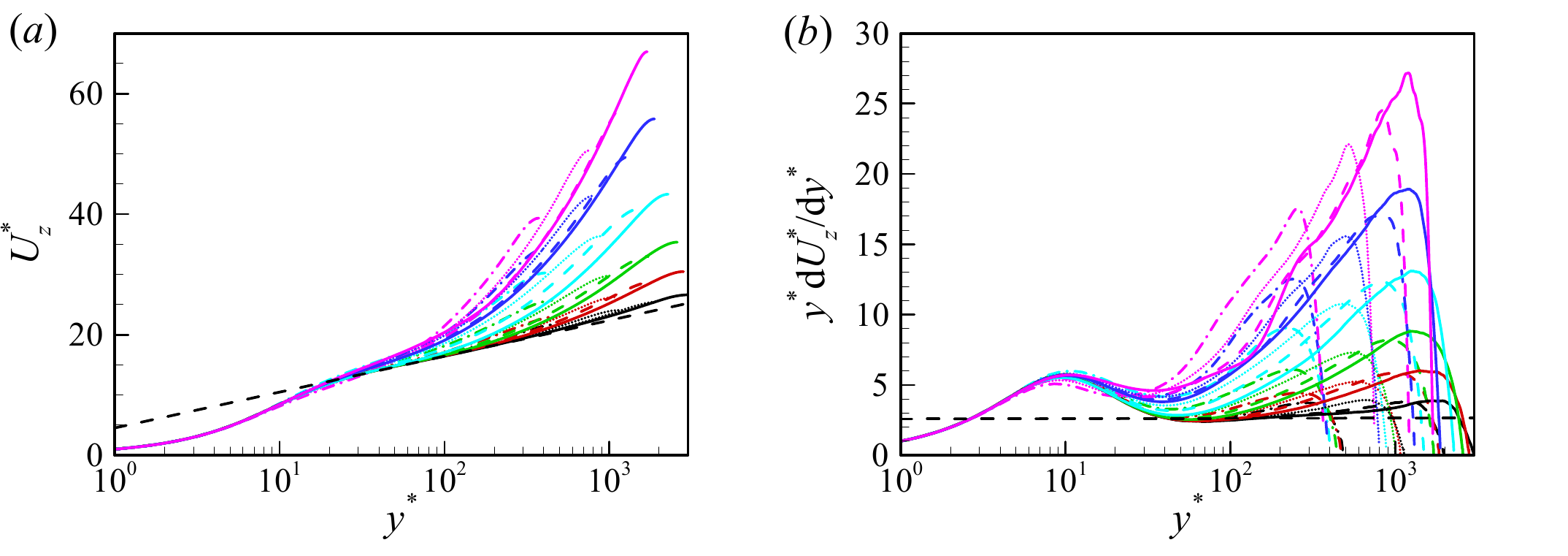}
  \caption{Inner-scaled mean axial velocity profiles (a), and corresponding logarithmic diagnostic functions (b). 
	The colour codes correspond to the same values of the rotation number:
	$N=0$ (black), 
	$N=0.25$ (red), 
	$N=0.5$ (green), 
	$N=1$ (cyan), 
	$N=2$ (blue), 
	$N=4$ (pink). 
	The line patterns correspond to the same values of the nominal Reynolds number:
	$\Rey_{\tau,0}=495$ (dash-dotted),
	$\Rey_{\tau,0}=1137$ (dotted),
	$\Rey_{\tau,0}=1979$ (dashed),
	$\Rey_{\tau,0}=3020$ (solid).
        The dashed black lines in panel (a) denote the expected 
	logarithmic distribution given in equation~\eqref{eq:uplus},
	and in panel (b) the inverse of the Karman constant ($1/\kappa$).)
	}
  \label{fig:uplus}
\end{figure}

To analyze the flow modifications yielding drag reduction in greater detail, 
in figure~\ref{fig:uplus}(a) we display the inner-scaled mean axial velocity 
profiles at various rotation and Reynolds numbers. At zero and small rotation numbers, 
the mean axial velocity follows with good accuracy the 
logarithmic distribution observed in pipe flow~\citep{Pirozzoli2021b},
\begin{equation} 
U_z^* = \frac{1}{\kappa} \log y^* + B, \label{eq:uplus} 
\end{equation}
with $\kappa = 0.387$ and $B = 4.53$. Deviations 
from such universal behaviour occur farther and farther from the wall as the Reynolds number increases. 
However, deviations tend to occur earlier as $N$ increases, and at the same time, 
the wake is found to grow stronger. At $N \gtrsim 1$, the whole logarithmic layer 
is eventually disrupted, consistent with the findings of \citet{Orlandi1997a, Davis2019}.
This process is clearer when the logarithmic diagnostic function, namely $y^* \mathrm{d}U_z^*/\mathrm{d}y^*$, 
is considered, as shown in figure \ref{fig:uplus}(b).
Indeed, no plateau of this indicator is found at $N \gtrsim 1$, although one can speculate that 
a logarithmic behaviour is recovered at higher Reynolds numbers than we consider here.

\begin{figure}
  \centering
  \includegraphics[width=\linewidth]{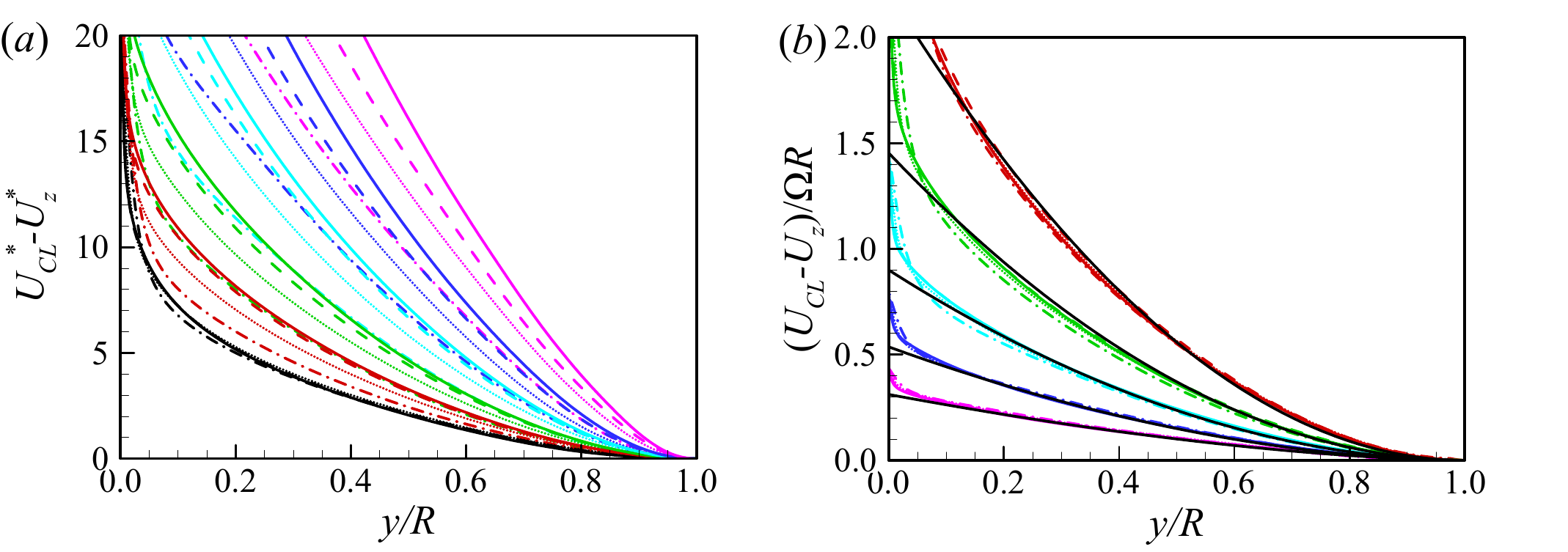}
  \caption{Mean axial velocity defect profiles ($U_{CL}$ is the pipe centreline velocity), 
	normalized by the friction velocity, $u_{\tau}$ (a), 
	and by the pipe rotational velocity, $\Omega R$ (b).
	The colour codes correspond to the same values of rotation number:
	$N=0$ (black), 
	$N=0.25$ (red), 
	$N=0.5$ (green), 
	$N=1$ (cyan), 
	$N=2$ (green), 
	$N=4$ (pink). 
	The line patterns correspond to the same values of the nominal Reynolds number:
	$\Rey_{\tau,0}=495$ (dash-dotted),
	$\Rey_{\tau,0}=1137$ (dotted),
	$\Rey_{\tau,0}=1979$ (dashed),
	$\Rey_{\tau,0}=3020$ (solid).
	The solid black lines in panel (b) indicate the power-law fitting function \eqref{eq:defect},
	with coefficients given in \eqref{eq:fit}.
	}
  \label{fig:udefect}
\end{figure}

The mean axial velocity profiles are then shown in defect form in figure~\ref{fig:udefect}. 
As known from previous studies \citep{Pirozzoli2021b}, normalization by the friction 
velocity (panel (a)) yields excellent universality of the profiles in the absence of 
rotation ($N=0$). However, when rotation is introduced, a large scatter is observed when 
either the rotation number or the Reynolds number vary. Indeed, as noticed 
by \citet{Oberlack1999}, the pipe rotational speed is an independent velocity scale, 
and it is a better candidate to achieve universality of the mean axial velocity profiles
in the presence of pipe rotation.
Hence, in panel (b) we show the velocity profiles normalized by the peripheral rotational speed. 
Excellent universality of the axial velocity profiles is then observed for any given 
(non-zero) rotation number, however with obvious dependence on $N$. Of course, 
this kind of normalization cannot apply to the non-rotating case as $\Omega=0$. 
Based on these empirical observations, we then assume the following form for the defect velocity profiles,
\begin{equation}
\frac{U_{CL} - U_z}{\Omega R} = \frac{1}{N} \varphi(N) \left(\frac{r}{R}\right)^{\alpha(N)}, \label{eq:defect}
\end{equation}
where fitting of the DNS data in the range $0 \leq r/R \leq 0.9$ yields
\begin{equation}
\varphi(N) = 0.90 + 0.25 \log N, \quad \alpha(N) = 2.0 - 0.071 N^{1.2}. \label{eq:fit}
\end{equation}
It should be noted that previous experiments \citep{Kikuyama1983} and DNS \citep{Orlandi1997a} 
also suggested values of the power-law exponent $\alpha \approx 2$, however, based on a much more limited set of data.

The previous observations can be leveraged to obtain predictions for the friction coefficient. 
Indeed, integration of the defect profile~\eqref{eq:defect} yields the following relationship 
between the centreline and bulk velocity 
\begin{equation}
U_{CL}^* = u_b^* \left( 1 + \frac{2}{2 + \alpha} \varphi \right). \label{eq:Uclp}
\end{equation}
Matching the wall-normal gradients of the inner profile \eqref{eq:uplus} with the defect profile \eqref{eq:defect} 
yields the condition 
\begin{equation}
\frac{\mathrm{d} U_z^*}{\mathrm{d} \eta} = \frac{1}{\kappa \eta} = \alpha \varphi u_b^* \left( 1 - \eta \right)^{\alpha - 1}, \label{eq:match1}
\end{equation}
where $\eta=y/R$. This condition can be numerically solved to determine the outer-scaled matching location $\eta_0$. Under the assumption $\eta_0 << 1$, the approximate solution holds 
\begin{equation}
\eta_0 \approx \frac{1}{\kappa \alpha \varphi u_b^*}. \label{eq:etas}
\end{equation}
Finally, matching the pointwise values of \eqref{eq:uplus} and \eqref{eq:defect} at $\eta = \eta_0$ yields 
\begin{equation}
\frac{1}{\kappa} \log \eta_0 + \frac{1}{\kappa} \log \Rey_{\tau} + B = u_b^* \left( 1 + \frac{2 \varphi}{2 + \alpha} - \varphi \left( 1 - \eta_0 \right)^{\alpha} \right). \label{eq:match2}
\end{equation}
Equation~\eqref{eq:match2} can be numerically solved to obtain $u_b^*$ for a given $\Rey_b$ (as $\Rey_{\tau} = \Rey_b / (2 u_b^*)$) and $N$, and in turn obtain the friction factor, $f = 8 / {u_b^*}^2$.

\begin{figure}
  \centering
  \includegraphics[width=\linewidth]{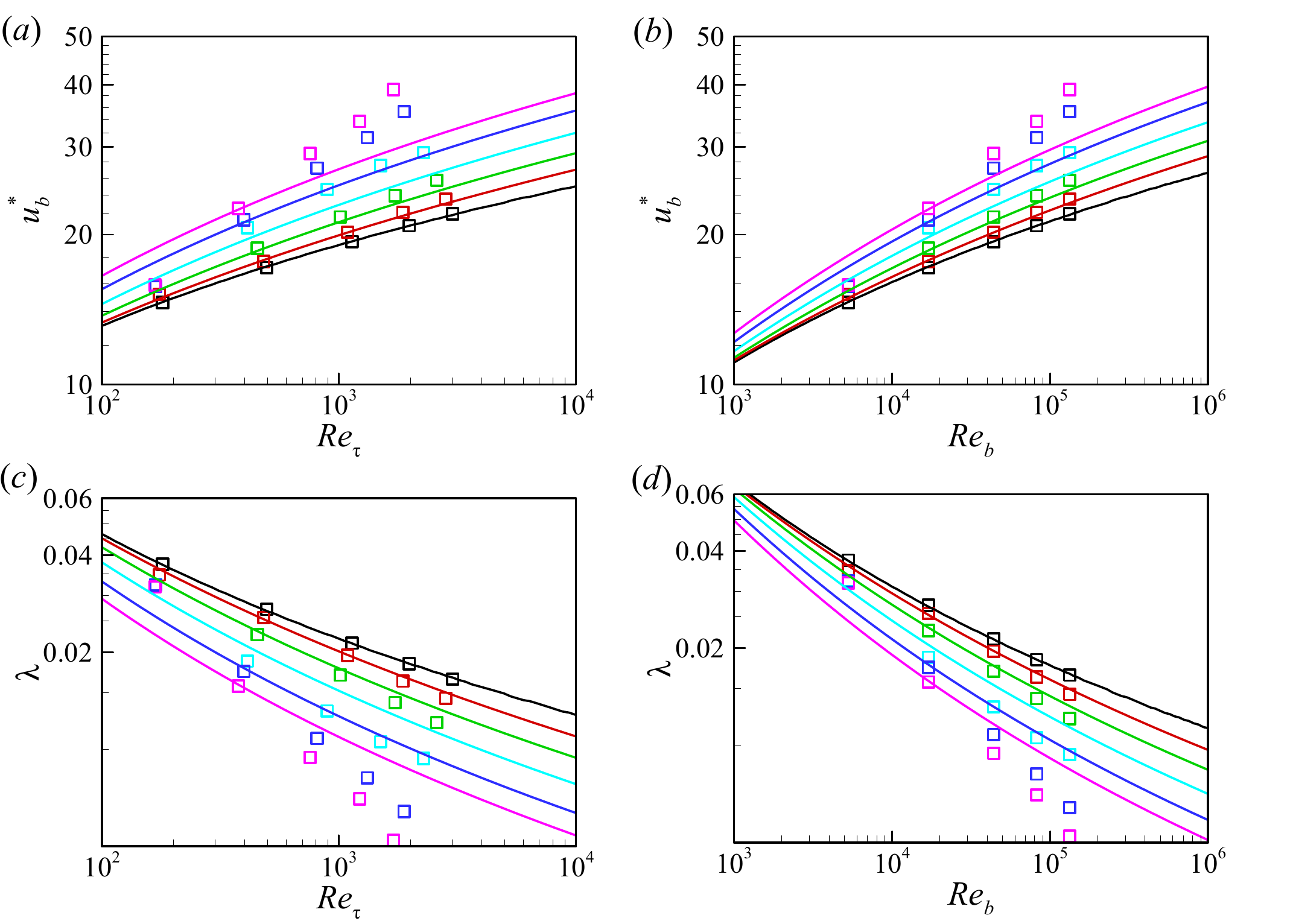}
	\caption{Inner-scaled bulk velocity (a,b) and friction factor (c,d), as a 
	function of the friction Reynolds number (a,c) and bulk Reynolds number (b,d).
	The colour codes correspond to the same values of rotation number:
	$N=0$ (black), 
	$N=0.25$ (red), 
	$N=0.5$ (green), 
	$N=1$ (cyan), 
	$N=2$ (blue), 
	$N=4$ (pink). 
	Symbols denote the DNS data, and lines the corresponding predictions of equation~\eqref{eq:match2}.
	The solid black line denotes the predictions of Prandtl's friction law for a non-rotating pipe~\citep{Pirozzoli2021b}.
	}
  \label{fig:ubp}
\end{figure}

The charts of $u_b^*$ and $f$ as a function of the Reynolds number are shown in figure~\ref{fig:ubp}, 
where symbols denote the DNS data, and lines represent the corresponding predictions of 
equation~\eqref{eq:match2}. 
The latter formula indeed captures the correct Reynolds number trends 
including the progressive departure of the friction curves from the 
non-rotating case, and it yields good quantitative predictions
at low-to-moderate rotation numbers. 
However, the drag reduction effect becomes noticeably underestimated at $N \gtrsim 1$, the reason being the 
previously noted breakdown of the logarithmic profile~\eqref{eq:uplus} at high rotation numbers, which we used for theoretical inference.
\begin{figure}
  \centering
  \includegraphics[width=\linewidth]{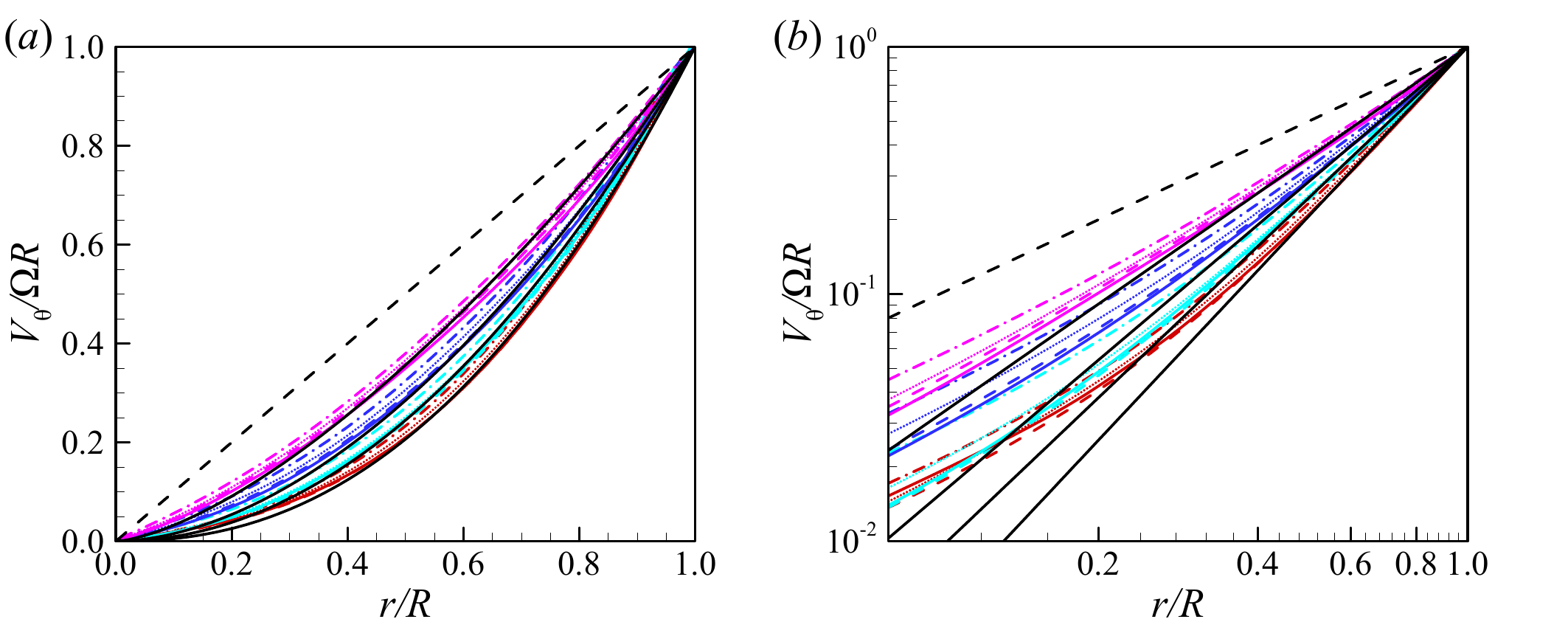}
	\caption{Radial profiles of mean azimuthal velocity in linear (a) and logarithmic (b) scale. The colour codes correspond to the same values of rotation number: $N=0.25$ (red), $1.0$ (cyan), $2.0$ (blue) and $4.0$ (pink). The line patterns correspond to the same values of the nominal Reynolds number: $\Rey_{\tau,0}=495$ (dash-dotted), $\Rey_{\tau,0}=1137$ (dotted), $\Rey_{\tau,0}=1979$ (dashed), $\Rey_{\tau,0}=3020$ (solid). The solid black lines denote equation \eqref{eq:vel_theta}, with the coefficient given in \eqref{eq:psi}. The black dashed lines correspond to the case of solid body rotation.}
  \label{fig:vel_theta}
\end{figure}

In figure~\ref{fig:vel_theta}, we further present the 
distributions of the absolute mean velocity in the azimuthal direction. Before analyzing its scaling laws, it is informative to consider the momentum balance equation projected along the $\theta$ direction,
\begin{equation}\label{eq:mom}
\nu \frac{\diff}{\diff r} \left( {\frac{1}{r}\frac{\diff}{\diff r} \left( {r V_\theta } \right)} \right) = \frac{1}{{{r^2}}}\frac{\diff}{{\diff r}}\left( {{r^2}\langle {{u_r}{u_\theta }} \rangle } \right).
\end{equation}
Multiplying \eqref{eq:mom} by $r^2$, then integrating from $0$ to $r$ yields
\begin{equation}\label{eq:urut}
	\nu r \frac{\diff}{\diff r}\left( \frac{V_\theta} {r} \right) = \left\langle {{u_r}{u_\theta }} \right\rangle,
\end{equation}
which highlights the balance between the viscous and the turbulent shear stress.
At the wall, $\langle u_r u_\theta \rangle = 0$, 
resulting in $V_\theta \overset{r\to R}{\operatorname*{\sim}}\Omega r$. 
As one moves away from the wall, the presence of a nonlinear distribution of $V_\theta$ 
becomes evident, due to positive values of $\left\langle {{u}_{r}}{{u}_{\theta }} \right\rangle$. \citet{Oberlack1999} suggested that, for $r/R \gtrsim 0.2$, the mean azimuthal velocity should follow a power-law variation, namely
\begin{equation}\label{eq:vel_theta}
    \frac{V_\theta}{\Omega R} = \left(\frac{r}{R}\right)^\psi,
\end{equation}
with $\psi = 2.0$, as inferred from the set of data available at that time, which according to \eqref{eq:urut} would result in linear variation of the
turbulent shear stress $\left\langle {{u}_{r}}{{u}_{\theta }} \right\rangle$, with slope proportional to $N$. 
However, \citet{Orlandi1997a} reported that the slope of $\langle u_r u_\theta \rangle$ 
is not strictly proportional to $N$, even in the core region, thereby preventing 
perfect collapse of the profiles in figure~\ref{fig:vel_theta}. 
In our simulations, $V_\theta/\Omega R$ is found to gradually increase with $N$, exhibiting 
a similar trend as other DNS \citep{Orlandi1997a,Davis2019,Feiz2005} and experimental 
studies \citep{Kikuyama1983}. We have checked the influence of pipe length on the 
azimuthal velocity profile, and confirmed that using $L=30R$ yields almost the same 
results as using $L=15R$ (see figure~\ref{fig:sens}(b)). It is worth mentioning that 
the experimental data of \citet{Reich1989} for $\Rey_b=5000$ with $N=1.0$ and $5.0$ 
are in good agreement with the quadratic law, however, they are likely affected by 
entrance effects, as pointed out by~\citet{Orlandi1997a}. Instead, our DNS data show that the exponent $\psi$ is clearly dependent on the rotation number, and also has weaker dependence on the bulk Reynolds number (panel (a)). As the rotation number increases, the azimuthal velocity profiles become closer to the case of solid body rotation, corresponding to smaller values of $\psi$. This tendency is regarded to be reasonable, considering that as the rotation number increases, turbulent motions become increasingly suppressed due to the stronger centrifugal instability. The weak influence of the nominal bulk Reynolds number includes slightly higher values of $\psi$ as $\Rey_b$ increases. This tendency is partly explained in terms of diminished viscous effects near the wall, which shortens the layer with approximately linear variation. 
Fitting the DNS data in the range $0.2 \leq r/R \leq 0.9$, we get
\begin{equation}\label{eq:psi}
    \psi(N) = 2.48-0.45N^{0.57},
\end{equation}
which is quite accurate as shown in figure~\ref{fig:vel_theta}. At $N \approx 1$, the formula yields $\psi \approx 2.0$, corresponding to a quadratic power law. We would like to note that for $r/R \lesssim 0.2$, the azimuthal velocity profiles observed in the DNS exhibit deviations from the quadratic law and become closer to a linear distribution (see panel (b)), thus indicating that the flow near the pipe axis is close to a solid-body rotation state.

\subsection{Organisation of turbulence} \label{sec:str}

\begin{figure}
  \centering
  \includegraphics[width=\linewidth]{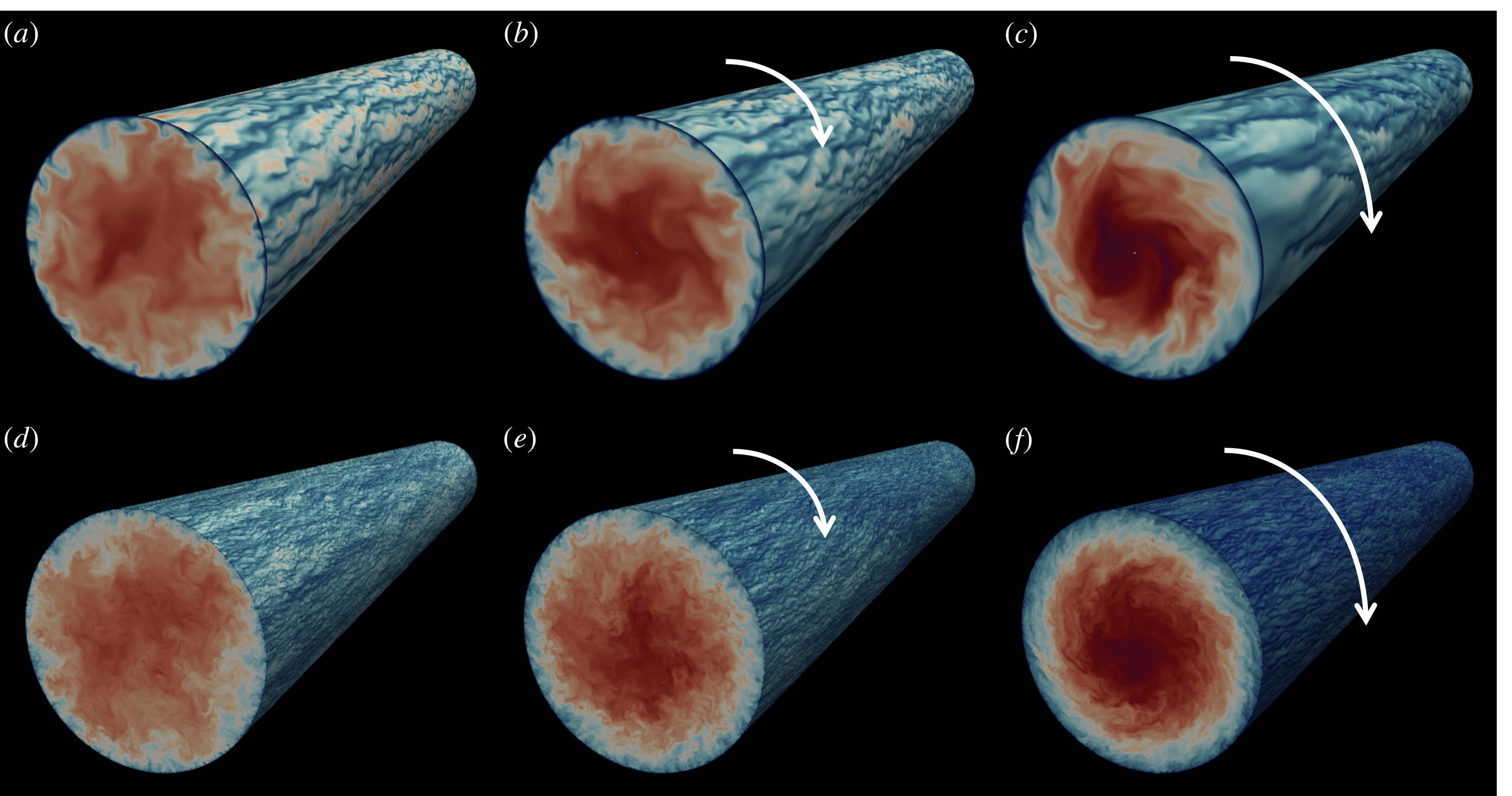}
	\caption{Instantaneous fields of axial velocity on a near-wall cylindrical shell and cross-stream plane. The cylindrical shell is located at a wall distance $y^+ \approx 15$. The colour scale is from blue (low speed) to red (high speed). Rotation numbers $N = 0, 0.5$ and $2.0$ are reported in panels (a,d), (b,e), (c,f), respectively, for $\Rey_{\tau,0} = 495$ (a-c) and $\Rey_{\tau,0} = 3020$ (d-f). The arrows indicate the direction of rotation.}
  \label{fig:str}
\end{figure}

A perspective view of the instantaneous axial velocity field is provided in figure~\ref{fig:str} for two selected Reynolds numbers, including for reference the case of pipe flow in the absence of rotation~\citep{Pirozzoli2021b}. As the Reynolds number increases, finer scales are observed, but the flow in the cross-stream plane is always dominated by a limited number of bulges distributed along the azimuthal direction, which correspond to alternating intrusions of high-speed fluid from the pipe core and ejections of low-speed fluid from the wall. Streaks are clear in the near-wall cylindrical shell, whose pattern has a close association with the turbulence organisation in the cross-stream plane. The $R$-sized low-speed streaks are linked to the large-scale ejections, and $R$-sized high-speed streaks are associated with the large-scale inrush from the core flow. Simultaneously, smaller streaks scaled in wall units prevail very close to the wall, which is correlated with buffer-layer ejections and sweeps. Hence, the organisation of the flow at two characteristic length scales is apparent, whose separation increases with the Reynolds number.


\begin{figure}
  \centering
  \includegraphics[width=\linewidth]{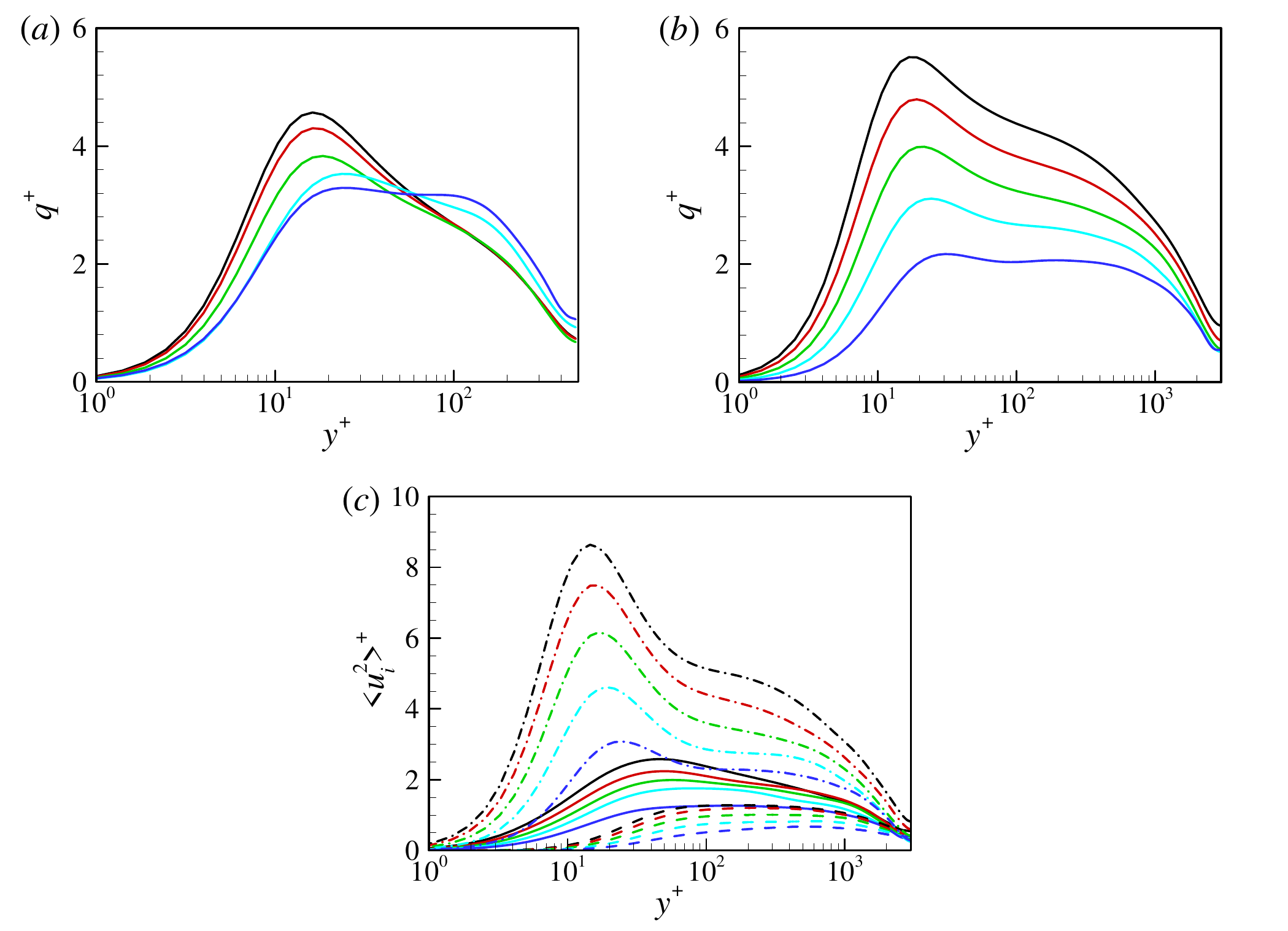}
	\caption{Distributions of: (a) turbulence kinetic energy for $\Rey_{\tau,0}=495$; (b) turbulence kinetic energy for $\Rey_{\tau,0}=3020$; (c) individual velocity variances for $\Rey_{\tau,0}=3020$. In (c), lines correspond to: $\langle u_\theta^2 \rangle$ (solid); $\langle u_r^2 \rangle$ (dashed); $\langle u_z^2\rangle$ (dash-dot). Color codes: $N = 0$ (black), $0.25$ (red), $0.5$ (green), $1.0$ (cyan) and $2.0$ (blue).}
  \label{fig:tke}
\end{figure}

When rotation is introduced, a distinctive axial coherence becomes apparent in the core region~\citep{White1963}, reminiscent of the columnar structures observed in rotating homogeneous turbulence~\citep{Hu2022,Godeferd2015}. Meanwhile, the $R$-sized structures in the cross-stream plane undergo deformation due to the decreasing angular velocity towards the core. The bulges gradually tend to lag behind the azimuthal motion of the wall as one moves away from it. For the low-$\Rey$ case, the turbulent fluctuations are intensified in the core region for $N \gtrsim 1.0$, as highlighted by increased turbulence kinetic energy, $q = \langle u_r^2 + u_\theta^2 + u_z^2 \rangle / 2$, in figure~\ref{fig:tke}(a).
\citet{Orlandi1997a} conjectured that such increase is due to a low-$\Rey$ effect, whereby the fluctuations in the central part of the pipe are amplified by the near-wall enlarged vortical structures. This conjecture is corroborated herein as the phenomenon is much less evident in the high-$\Rey$ case, as substantiated by decreased turbulence kinetic energy shown in figure \ref{fig:tke}(b). It is nevertheless important to highlight that at sufficiently low rotation numbers the fluctuations in the core region become consistently suppressed regardless of the Reynolds number. Turning to the near-wall region, a noticeable reduction in both the number and magnitude of the small-scale streaks is evident at both Reynolds numbers, which points to attenuation of sweeps and ejections in the buffer layer. This causes a distinct decrease in the axial velocity fluctuations compared with the other two components (see figure~\ref{fig:tke}(c)), which is not unexpected as the production of $q$ is associated with the axial velocity component. Moreover, the Taylor-Proudman theorem~\citep{Greenspan1968} implies that the vorticity component normal to the axis of rotation should be suppressed to reduce the axial velocity gradients. Noteworthy is the lack of discernible azimuthal tilting of the small-scale streaks. However, tilting of the footprints of the large-scale structures towards the direction of rotation is evident on the cylindrical shell in the low-$\Rey$ case. 

\begin{figure}
  \centering
      \includegraphics[width=\linewidth]{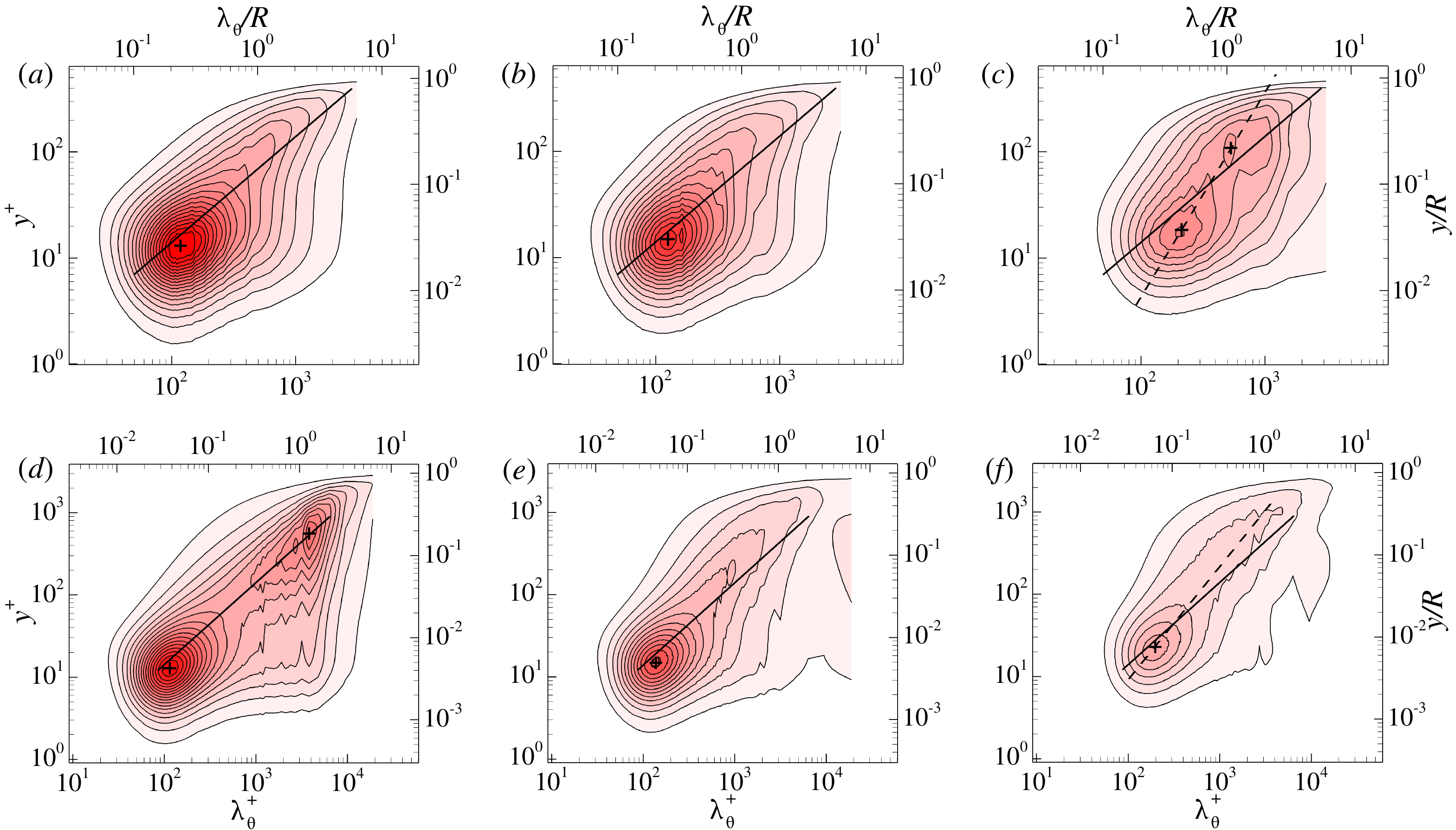}
  \caption{Pre-multiplied azimuthal energy spectra of axial velocity $\left( \kappa_\theta E_{uu}\right)^+$ for 
	$N=0$ (a, d), $N=0.5$ (b, e), $N=2.0$ (c, f), at 
	$\Rey_{\tau,0} = 495$ (a-c), 
	$\Rey_{\tau,0} = 3020$ (d-f). 
	The solid lines denote the trend $y = 0.14 \lambda_\theta$, and the dashed lines denote the trend $\lambda_\theta/R = 2.35(y/R)^{0.52}$ in panel (c), and $\lambda_\theta/R = 2.20(y/R)^{0.72}$ in panel (f).
	The crosses denote spectral peaks.}
  \label{fig:spec}
\end{figure}

The modifications in the characteristic length scales can be quantified in terms of the 
pre-multiplied azimuthal energy spectra of the axial velocity, which we report in figure~\ref{fig:spec}.
In the non-rotating cases, the spectra exhibit a prominent ridge along $\lambda_\theta \sim y$, which should be interpreted as a hierarchy of wall-attached eddies as suggested by \citet{Townsend1976}. Here, $\lambda_\theta$ is the wavelength in the azimuthal direction. The inner peak, associated with the near-wall turbulence regeneration cycle~\citep{Jimenez1999}, and the outer site, linked with $R$-sized large-scale motions~\citep{Hutchins2007}, exhibit a more pronounced separation in the high-$\Rey$ case. Upon imposition of rotation, the most remarkable modification is the progressive attenuation of the amplitude of the inner spectra. Concurrently, an outer-layer peak of $\lambda_\theta/R=1.08$ emerges at $y/R=0.22$ for $N=2.0$ in the low-$\Rey$ case, which we believe to be linked with the $R$-sized distorted structures observed in figure~\ref{fig:str}(c). In contrast, the outer-layer peak of $\lambda_\theta/R=1.26$ at $y/R=0.19$ in the high-$\Rey$, non-rotating case vanishes, and a subdominant peak is found instead at $\lambda_\theta/R=\pi$, at a wall distance $y^+ \lesssim 200$. At both Reynolds numbers, the inner peak undergoes a top-right shift, from about $y^+=13$ to $19$, with a concurrent increase in the typical streaks spacing from about $\lambda _{\theta }^{+}=120$ to about $200$. Consequently, the spectral ridge becomes steeper, taking the form of a power law, with the exponent changing with $\Rey$ and $N$.


\subsection{Contributions to frictional drag}  \label{sec:fik}

The Fukagata-Iwamoto-Kasagi (FIK) identity~\citep{Fukagata2002} is herein used to isolate the contributions of molecular viscosity and turbulence to the overall wall friction, and to elucidate mechanisms contributing to the observed drag reduction. 
It can be shown that the form of the FIK identity for a rotating pipe is identical to the case of a non-rotating pipe as reported by \citet{Fukagata2002}, because its derivation originates from the momentum equation in the $z$ direction, which does not depend on the imposition of axial rotation. This results in the following expression for the relative drag reduction,
\begin{equation}\label{eq:dr_fik}
	\DR = 1 - \frac{C_f}{C_{f,0}} = \int_{0}^{\log \Rey_{\tau,0}} 
	\left( \frac yR \right) \left( {c_f^T}_0(y^+) - {c_f^T}(y^+) \right) \diff \log y^+,
\end{equation}
where 
\begin{equation}\label{eq:cf}
	{c_f^T}(y^+) = 4 \left(\frac{r}{R}\right)^2 \langle u_r u_z \rangle^+,
\end{equation}
denotes the local contribution of turbulence at a given wall distance to the overall friction coefficient.
It is noteworthy that the viscous contribution to wall friction ($16/\Rey_b$) is the same at the various rotation numbers as the DNS were carried out at constant mass flow rate, hence it cancels out from \eqref{eq:dr_fik}.
\begin{figure}
  \centering
      \includegraphics[width=\linewidth]{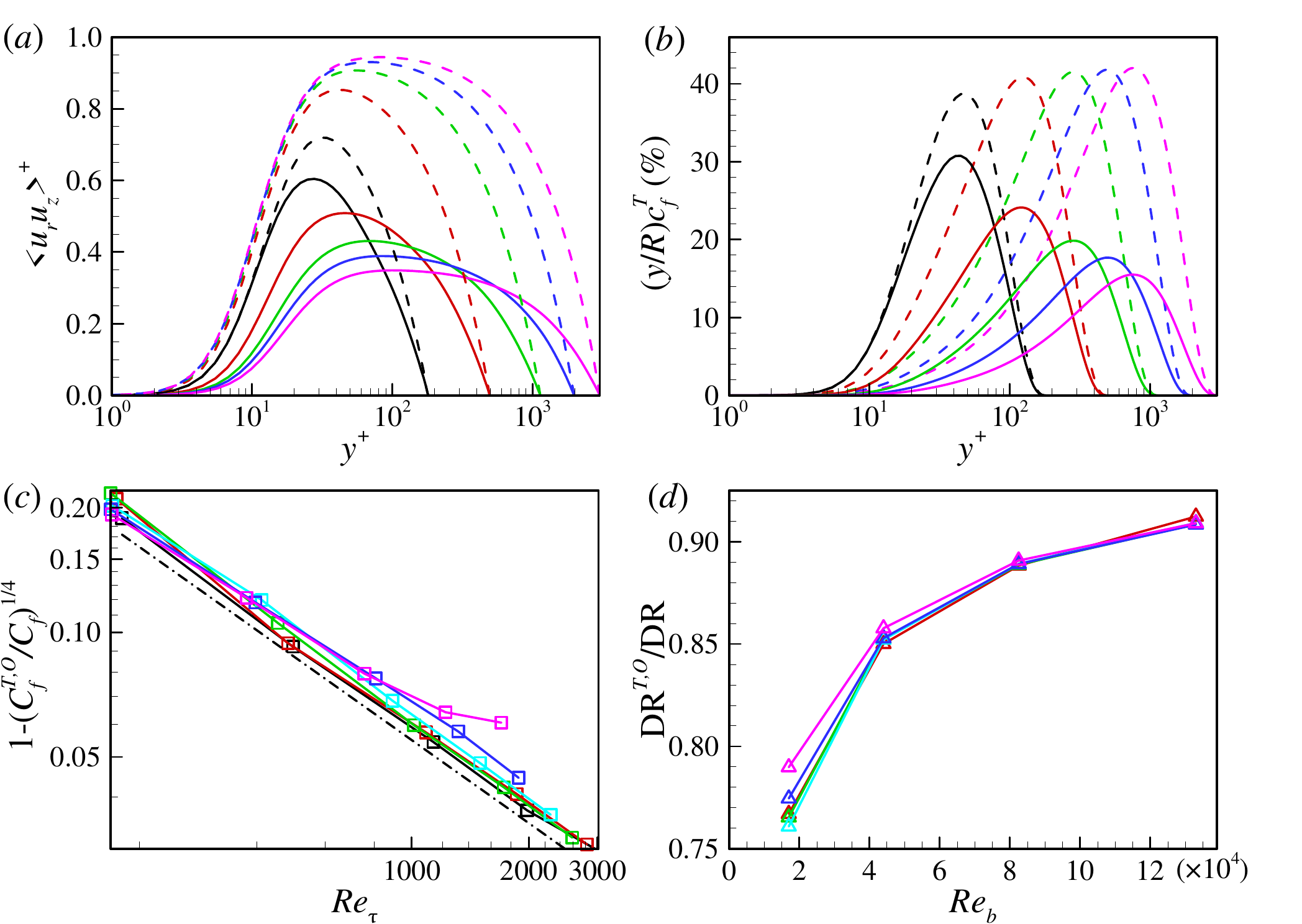}
	\caption{(a) Turbulent shear stress for $N=0$ (dashed lines) and $N=2$ (solid lines); (b) local contributions to turbulent friction (as from equation~\eqref{eq:cf}), for $N=0$ (dashed lines) and $N=2$ (solid lines); (c) turbulent contribution to friction as a function of $\Rey_\tau$; (d) fraction of drag reduction due to turbulence as a function of $\Rey_b$. In (a) and (b), $\Rey_{\tau,0} = 180$ (black), $495$ (red), $1137$ (green), $1979$ (blue) and $3020$ (pink). In (c) and (d), $N=0$ (black), $0.25$ (red), $0.5$ (green), $1.0$ (cyan), $2.0$ (blue) and $4.0$ (pink). In (c), the dash-dotted line indicates the prediction of equation~\eqref{eq:cfto_ulti}.}
  \label{fig:fik}
\end{figure}
We have made sure that the relative errors between friction coefficients obtained from the direct calculation and the FIK identity are well below 1\%.

Figure~\ref{fig:fik}(a) clearly shows that the reduction of the turbulent shear stress is greater at higher Reynolds numbers. Whereas the maximum shear stress slowly increases asymptotically to unity as $\Rey_b$ increases, the peak decreases in the case of a rotating pipe, despite amplification of turbulence kinetic energy in the core region at high rotation numbers in the low-$\Rey$ cases (see figure~\ref{fig:tke}(a)).
This can be interpreted better by noticing that the inner-scaled wall rotational speed, $N^*=N/2(\Rey_b/\Rey_{\tau})$, increases with $\Rey_b$, with likely increased suppression of the near-wall turbulence.
Figure~\ref{fig:fik}(b) reports the wall-normal distribution of the local turbulence contribution to the overall friction.
Note that the plot is reported in a semi-log scale and in pre-multiplied form, to have a correct perception of the integrated contributions. 
The figure shows that, at low Reynolds numbers, suppression of the turbulent contribution to friction
mainly takes place in the core part of the flow, with a reduction of the absolute peak.
At higher Reynolds numbers, a strong decrease of the peak value is still present and increasing, but substantial drag reduction is also coming from the near-wall layer.
To have a quantification of the effects of the inner and outer layers,
we then split the turbulent wall friction into integrated contributions
from the inner layer (say, $C_f^{T,I}$) and from the outer layer (say, $C_f^{T,O}$),
based on the position of the peak turbulent shear stress (say, $y_m$), as
suggested by \citet{Hurst2014}. Accordingly, we define
\begin{equation}\label{eq:cfto}
	\frac{C_{f}^{T,O}}{{{C}_{f,0}}} = \int_{\log y_m^+}^{\log \Rey_{\tau,0}} \left( \frac yR \right) {c_f^T}(y^+) \, \diff \log y^+.
\end{equation}
In non-rotating pipe flow, the turbulent shear stress 
attains a peak at $y_{m}^{+}\simeq c{{\left( \Rey_\tau/\kappa \right)}^{n}}$ (\citet{Chen2018} recommended $n=1/3$ for $\Rey_{\tau} \le 3000$),
and farther from the wall it decays linearly, hence $\langle u_r u_z \rangle ^+ \sim r/R$.
Under these assumptions, equation~\eqref{eq:cfto} yields
\begin{equation}\label{eq:cfto_ulti}
1-\left( \frac{C_f^{T,O}}{C_f}\right )^{1/4} \approx 4.0 (\kappa \Rey_\tau ^ 2 ) ^{-1/3}.
\end{equation}
Figure~\ref{fig:fik}(c) shows that \eqref{eq:cfto_ulti} also yields a satisfactory approximation in the case of rotating pipes, 
as long as $N \leq 2$. 
The formula also well underscores that $C_{f}^{T,O}$ asymptotically approaches $C_{f}$ with increasing $\Rey_\tau$, 
as a result of the increasing influence of the very large turbulent structures that populate the outer layer 
and modulate the small near-wall structures~\citep{Hutchins2007}.
In equation~\eqref{eq:cfto_ulti}, the splitting location for the rotating cases is based on the
respective profiles of $\langle u_r u_z \rangle$, but the asymptotic behaviour remains valid if $y_m$ 
is determined based on the non-rotating cases, since $y_{m,0} \lesssim y_m$, as inferred from figure \ref{fig:fik}(a). 
In figure~\ref{fig:fik}(d), we present the ratio of the drag reduction associated with the outer layer 
to the total drag reduction. 
This indicator is observed to exceed 90\% for $N=4.0$ and $\Rey_{\tau,0} = 3020$,
tending asymptotically to unity, consistent with theoretical expectations. 
We also emphasize that it becomes almost independent from $N$ at high enough Reynolds numbers. 
It is important to note that the FIK identity per se does not imply any direct causality link between its right- and left-hand sides, meaning that the observed reduction of the turbulent shear stress in the outer layer could as well be a consequence of the reduced wall friction, rather than the opposite.

\section{Conclusions}\label{sec:conclusion}

Direct numerical simulations have been performed to study axially rotating pipe flow up to nominal friction Reynolds number $\Rey_{\tau,0} = 3020$. The drag reduction rate, which we show to be equivalent to net power saving in the fully-developed scenario assuming no mechanical losses, increases with either $N$ and $\Rey_{\tau,0}$, becoming a sole function of the inner-scaled wall rotational speed at high enough Reynolds numbers. The drag reduction is as high as about 70\% at $N=4.0$ for $\Rey_{\tau,0}=3020$, although the flow is still far from full relaminarisation. We have developed a theoretical analysis 
informed with DNS data which allows for quantitative prediction of this effect, based on the observation
that the rotational speed becomes the relevant velocity scale in the core flow. The analysis yields a predictive formula 
for the friction coefficient~\eqref{eq:match2}, which yields accurate approximation of the DNS data up to $N \approx 1$,
regardless of the bulk Reynolds number, which however breaks down once the logarithmic layer in the mean axial velocity profile is disrupted.

Analysis of the instantaneous velocity fields reveals the role of rotation in weakening the near-wall sweeps and ejections and in the elongation and broadening of the streaks. In the core region, rotation leads to turbulence suppression at moderate rotation numbers and high Reynolds numbers, resulting in the disappearance of the outer-layer peak in the pre-multiplied spectra. 
Turbulence is instead intensified at moderate rotation numbers and low Reynolds numbers, and a distinct outer-layer peak or plateau, emerges in the pre-multiplied spectra. Nevertheless, a consistent decrease of the turbulent shear stress is observed. Use of the FIK identity reveals that the turbulent drag reduction originating from the outer layer asymptotically approaches the total turbulent drag reduction with increasing $\Rey_{\tau,0}$, consistent with a theoretically derived formula, and conveys that both the inner and outer layers increasingly contribute to drag reduction as $N$ increases. 

It is finally important to acknowledge opportunities and challenges in applying wall rotation in practical contexts. 
On the positive side, it is clear from this paper that large drag reduction is possible without reverting to complicated wall actuation rules. 
In particular, we note that at a statistically steady state, the work required to sustain rotation is zero (leaving mechanical losses aside), 
and drag reduction effects increase with the Reynolds number, unlike in conventional wall actuation techniques. 
On the negative side, setting a full pipeline into rotation may not be an easy task. In addition, we find that the achievement of a fully developed state requires a longer distance than for non-rotating flows (typically, a few hundred diameters), and energy must be spent during this transient state to put the whole system into rotation. This energy expenditure should be accounted for in the 
evaluation of axial rotation for realistic applicative scenarios.

\backsection[Supplementary movies]{Supplementary movies are available at \\
\href{https://doi.org/10.1103/APS.DFD.2023.GFM.V0080}{https://doi.org/10.1103/APS.DFD.2023.GFM.V0080}.}

\backsection[Acknowledgements]{We acknowledge that the simulations were performed using the EuroHPC Research Infrastructure resource LEONARDO based at CINECA, Casalecchio di Reno, Italy, under a LEAP grant. Maochao Xiao would like to acknowledge the financial support provided by the China Scholarship Council.}

\backsection[Declaration of interests]{The authors report no conflict of interest.}

\backsection[Data availability statement]{DNS data are available at \href{http://newton.dma.uniroma1.it/}{http://newton.dma.uniroma1.it/}.}

\backsection[Author ORCIDs]{\\
Maochao Xiao~\href{https://orcid.org/0000-0001-6528-2641}{https://orcid.org/0000-0001-6528-2641}; \\
Alessandro Ceci~\href{https://orcid.org/0000-0001-6664-1677}{https://orcid.org/0000-0001-6664-1677}; \\
Paolo Orlandi~\href{https://orcid.org/0000-0002-0305-5723}{https://orcid.org/0000-0002-0305-5723}; \\
Sergio Pirozzoli~\href{https://orcid.org/0000-0002-7160-3023}{https://orcid.org/0000-0002-7160-3023}.}

\appendix
\section{Numerical issues}\label{app:A}
\begin{figure}
  \centering
      \includegraphics[width=\linewidth]{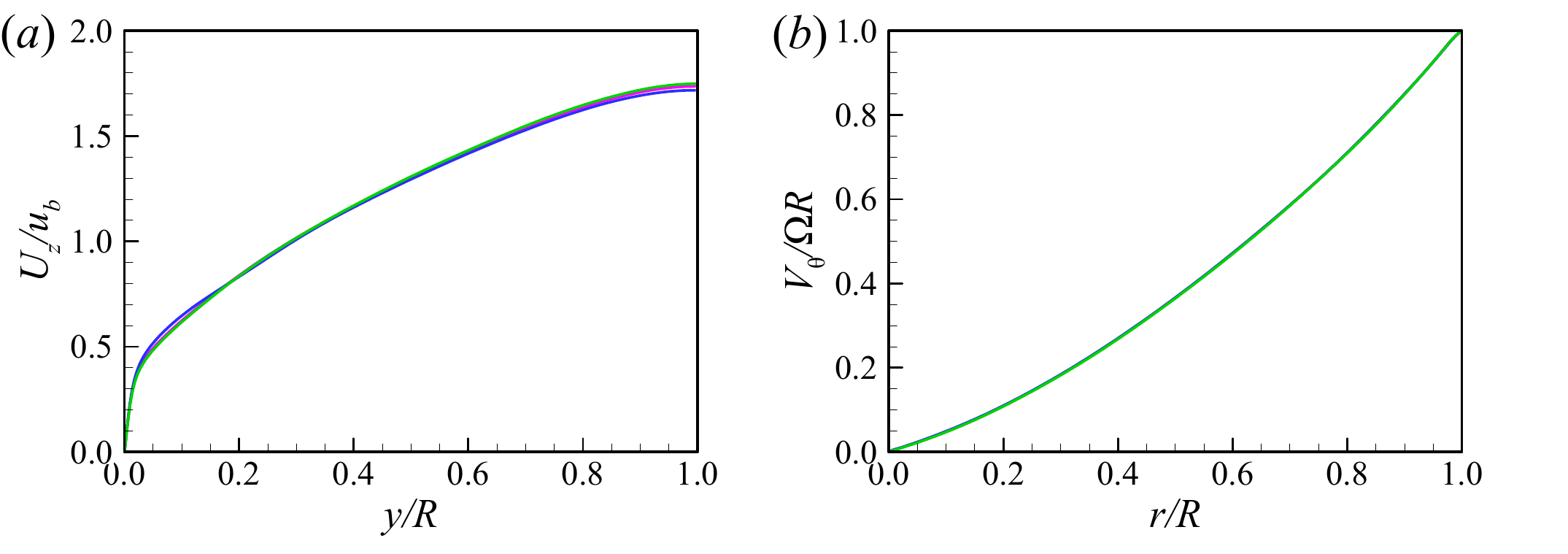}
  \caption{Radial profiles for $\Rey_{\tau,0} = 1137$ and $N=4.0$: (a) axial velocity; (b) azimuthal velocity. Pink: $L = 15R$ and $N_r=164$; blue: $L = 30R$ and $N_r=164$; green: $L=15R$ and $N_r=328$.}
  \label{fig:sens}
\end{figure}
The DNS herein reported assume a pipe length $L = 15R$. Figure~\ref{fig:sens} compares the profiles of axial velocity and azimuthal velocity obtained using $L=15R$ and $30R$ for the case of $\Rey_{\tau,0} = 1137$ and $N = 4.0$. The longer configuration predicts a more turbulent mean axial velocity profile. The friction factors for the two lengths have a relative difference of only about 3\%, which is much less than the drag reduction amount of about $55\%$ for this case (see Table \ref{tab:params}). Note that the discrepancies would be smaller with decreasing $N$ due to the weakened long columnar structures in the core region. Figure~\ref{fig:sens} also displays the results obtained using a grid with a higher resolution in the radial direction. The refined resolution results in a slightly less turbulent profile of the mean axial velocity, and it brings out a difference in the friction factor of only about $3\%$, which is again much lower than the drag reduction amount. Hence, we conclude that the current numerical settings could be considered enough especially when drag reduction is the primary focus.
 \begin{figure}
  \centering
      \includegraphics[width=0.45\linewidth]{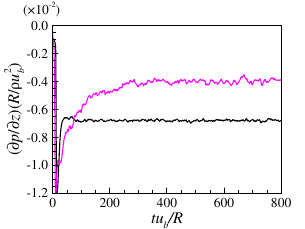}
  \caption{Time history of axial pressure gradient for $\Rey_{\tau,0} = 495$. Black: $N = 0$; pink: $N = 4.0$.} 
  \label{fig:dpdz}
\end{figure}
In our time-evolving simulations, the time interval needed to achieve a fully developed state becomes significantly longer with increasing $N$ at a fixed Reynolds number. Figure~\ref{fig:dpdz} shows the time history of the axial pressure gradient for $\Rey_{\tau,0} = 495$ at $N = 0$ and $4.0$, showing that it takes a time interval of about $100 R/u_b$ for the non-rotating case to become fully developed, whereas the rotating case needs about $400 R/u_b$. 

\bibliographystyle{jfm}
\bibliography{jfm}

\end{document}